\documentstyle[12pt,epsf]{article}
\def\beq{\begin{equation}}
\def\eeq{\end{equation}}

\def\beqn{\begin{eqnarray}}
\def\eeqn{\end{eqnarray}}
\def\ra{\rightarrow}
\def\prd#1#2#3{{\it Phys. Rev.} {\bf D#1} #2 (19#3)}

\def\np#1#2#3{{\it Nucl. Phys.} {\bf B#1} #2 (19#3)}
\def\prl#1#2#3{{\it Phys. Rev. Lett.} {\bf #1} #2 (19#3)}
\def\cp{{\cal CP}}
\def\pp{\overline{p}}
\def\uu{\overline{u}}

\def\nn{\overline{\nu}}
\def\goes{\stackrel{\cp}{\longleftrightarrow}}
\def\dd{\overline{d}}

\relax
\begin{document}
\begin{titlepage}
\def\ba{\begin{array}}
\def\ea{\end{array}}
\def\thefootnote{\fnsymbol{footnote}}
\vfill
\hskip 4in {BNL-61590}

\hskip 4in {ISU-HET-95-1}

\hskip 4in {March 1995}
\vspace{1 in}
\begin{center}
{\large \bf LOOKING FOR $\cp$ VIOLATION IN $W$ PRODUCTION AND DECAY}\\

\vspace{1 in}
{\bf S.~Dawson$^{(a)}$}\footnote{This manuscript has been authored
under contract number DE-AC02-76CH00016 with the U.S. Department
of Energy.  Accordingly, the
U.S. Government retains a non-exclusive, royalty-free license to
publish or reproduce the published form of this contribution, or
allow others to do so, for U.S. Government purposes.}
{\bf  and G.~Valencia$^{(b)}$}\\
{\it  $^{(a)}$ Physics Department,
               Brookhaven National Laboratory,  Upton, NY 11973}\\
{\it  $^{(b)}$ Department of Physics and Astronomy,
               Iowa State University,
               Ames IA 50011}\\
\vspace{1 in}
\end{center}
\begin{abstract}

We describe $\cp$ violating observables in resonant $W^\pm$ and $W^\pm$
plus one jet production at the Tevatron. We present simple examples of
$\cp$ violating effective operators, consistent with the symmetries of
the Standard Model, which would give rise to these observables. We find
that $\cp$ violating effects coming from new physics at the $TeV$ scale
could in principle be observable at the Tevatron with $10^6$ $W^\pm$ decays.

\end{abstract}

\end{titlepage}

\clearpage

\section{Introduction}

The origin of $\cp$ violation remains one of the unsolved questions in
particle physics. It is therefore very important to search for signals
of $\cp$ violation in all the experimentally accessible systems. The
Tevatron has now accumulated a sample of about $50 000$ $W^\pm$ events
and samples of $10^7$ events should be possible eventually. This makes
it timely to think of testing $\cp$ violation in $W^\pm$
production and decay.

In this paper we study the processes $p \pp \ra W^\pm$ and $p \pp \ra
W^\pm \ + 1 {\rm ~jet}$.\footnote{The Standard Model
perturbative amplitudes for $W$ production are given in
Ref. \cite{smprod}.  The re-summed amplitudes (valid at small transverse
momenta) are given in Ref. \cite{resumw}.}
 Some $\cp$ odd observables in these
processes involve the polarization of the $W^\pm$ boson. It is,
therefore, convenient to allow the $W^\pm$ to decay leptonically into
$\ell^\pm \nu$ pairs and study the complete processes
$p \pp \ra \ell^\pm \nu \ + \ {\rm ~0~or~1~jet}$. In this way we
recover some information on the direction of the vector boson polarization
by measuring the charged lepton momentum. Although it is also possible
to construct observables for hadronic $W^\pm$ decays, we will not
consider that case in this paper to avoid the complications of hadronization.

In this paper we adopt the strategy of searching
for small $\cp$ violating contributions to a dominant process instead
of looking for potentially larger effects in rare processes. We therefore
 limit our discussion to $\cp$ odd observables in the decay chains
\begin{equation}
p\pp \ra W^\pm \ + \ {\rm ~0~or~1~jet} \ra \ell^\pm \nu
\ + \ {\rm ~0~or~1~jet}
\end{equation}
and consequently we study these processes in the narrow width approximation.
We concentrate on the dominant parton subprocesses, ignoring for example,
top-quark distribution functions in the proton, $\cp$ violation in the parton
distribution functions, and $\cp$ violation that occurs only at higher twist.

It is, of course, possible that there are $\cp$ violating
effects that vanish within our approximation. An example in $\ell \nu$
production arises from the interference of the one-loop
electroweak corrections to the resonant
$W^\pm$ Standard Model amplitude with a non-resonant
$\cp$ violating four-fermion new physics
interaction. This example has been studied by Barbieri
{\it et. al} \cite{barbieri}.

It should be clear that the $\cp$ odd observables  that we discuss are
predominantly sensitive to $\cp$ violation in the $u d W$ vertex
and in the $W \ell \nu$ vertex  and
these are the same vertices which are probed in pion decay
$\pi \ra \ell \nu$.However, there are
several scenarios under which the direct $W$ production and decay
processes are more sensitive
to $\cp$ violation than the corresponding pion decay. This is true
in the examples we discuss for at least one of the following reasons:

\begin{itemize}

\item The new physics (at a high energy scale $\Lambda$)
that violates $\cp$ contributes to
the effective $W(q) f \overline{f^\prime}$ vertex in a way
proportional to $(q^2 /\Lambda^2)^n$, $n\geq 1$. Since
$q^2 \ll \Lambda^2$, the leading contribution will be for $n=1$.
The effects from the new physics on
direct $W$ production and decay are therefore
enhanced over effects in pion decay
by at least a factor $M_W^2 / m_\pi^2 \sim 3.5 \times 10^5$.

\item The absorptive phases needed to construct $T$-even
observables\footnote{By $T$-even (odd) observables we mean those
that do not change sign (do change sign) under the naive $T$ operation:
inversion of all momentum and spin vectors. Recall that this is not
the same as time reversal.}
are larger in $W$ production and decay. An example is the case where
the absorptive phase is due to a rescattering of the final state with
electroweak strength. This would be the case if the new physics generates
an effective four-fermion interaction that
contributes to the $W f \overline{f^\prime}$
vertex at the one-loop level. In this case, the absorptive phase introduces
an additional suppression factor of order
$\Gamma_W / M_W \sim 3\%$ for direct $W$ production and decay, whereas
it introduces a suppression factor of order $G_f m_\pi^2 \sim 10^{-8}$ for
pion decays.

\item Processes with a sufficient number of independent four-vectors
to construct $T$-odd triple products are suppressed at low energy.
In direct $W$ production and decay we can look for events with
one jet, suppressed by a factor of $\alpha_s$ with respect to the
lowest order  QCD process.
In pion decays we need a three body decay mode with the measurement of a
polarization or a four body decay mode. Three body decay modes are four
orders of magnitude smaller than $\pi \ra \ell \nu$ and polarization
measurements are extremely difficult. Four body decay modes are nine
orders of magnitude smaller than $\pi \ra \ell \nu$.

\end{itemize}

Kaon decay experiments tell us that the $\cp$ violating phases in the
Standard Model are extremely small. In addition, $\cp$ odd observables in
the standard model vanish in the limit of massless light fermions, and
are thus even smaller at higher energies \cite{gaso}.
The Standard Model does not
produce a sufficiently large $\cp$ violating signal to be observed in
the processes we study \cite{nacht}.
Popular extensions of the
standard model in the context of $\cp$ violation include multi-Higgs
models. In these models $\cp$ violation is also proportional to
fermion masses and thus negligible in processes like
 $d \uu \ra W^- \ra e^- \nn$
at high energy. We will, therefore, work from the assumption that
studies of $\cp$ violation at colliders
will only be sensitive to non-Standard Model sources.
Furthermore, we will not consider a specific model for $\cp$ violation,
but instead we will use an effective
Lagrangian approach to parameterize possible $\cp$ violating operators.
We will further assume that the origin of these operators lies in the
physics that breaks the electroweak symmetry. With a linear realization
of the symmetry breaking and a light Higgs, all the operators
of dimension $6$ have been given in Refs. \cite{buchmuller,burgess}.
With a non-linear realization of the symmetry breaking sector, anomalous
fermion-gauge-boson couplings have been described in Ref.~\cite{peccei}.
Here we consider just a few of these operators to illustrate the physics,
but it should be obvious that a similar analysis can be applied to other
operators.

There have been a number of studies of $\cp$ violation at colliders
that concentrate on effects due to the top-quark or multiple $W$ production
Ref.~\cite{rindani}. We concentrate on single $W$ production and complement
the work of Refs.\cite{barbieri,brand}.
We will not discuss detector issues at all, except to make the obvious
statement that the detector must be ``$\cp$-blind'' in order to carry out these
studies. We limit ourselves to offer a ``proof in
principle'' that $\cp$ violation could be observed in $W^\pm$ decays
at the Tevatron.

\section{$p\pp \ra W^\pm X\ra \ell^\pm \nu X$}

We assume that the proton and anti-proton beams are unpolarized, and that
the polarization of the lepton is not measured. In this case, it is only
possible to construct $T$-even observables for this reaction. Some $\cp$
odd observables have been listed in Ref.~\cite{nacht}. Here we discuss
a few simple observables of this type in the context of a $\cp$ violating
four-fermion interaction due to physics beyond the minimal standard model.

Under a $\cp$ transformation, the reaction $p(\vec{p})\pp(-\vec{p}) \ra
\ell^+(\vec{q}) \nu X$ transforms into
$p(\vec{p})\pp(-\vec{p})\ra \ell^-(-\vec{q}) \nn \overline{X} $. Here
we work in the $p \pp$ center of mass frame and denote by $\overline{X}$
the $\cp$ conjugate of $X$. Also, we have only considered those kinematic
variables that can be observed, in this case the momenta of the beam and
the lepton. It is conventional to take the $z$-axis as being the direction
of the proton beam, and to use rapidity and transverse momentum as the
kinematical variables. Recalling that the lepton rapidity is given by
(all variables in the $p \pp$ center of mass):
\begin{equation}
y_\ell = {1\over 2}\log\biggl({E_\ell + q_{z \ell} \over E_\ell - q_{z \ell}}
\biggr)
\end{equation}
and the lepton transverse momentum by:
\begin{equation}
p_{T\ell}  = |\vec{q}\sin\theta|
\end{equation}
where $\theta$ is the angle between the proton and lepton momenta in the lab
frame, one can see that under a $\cp$ transformation
\begin{equation}
y_{\ell^-} \goes -y_{\ell^+},\ \ \
p_{T\ell^-} \goes p_{T\ell^+}
\end{equation}
In terms of these variables we can, therefore, construct $\cp$-odd
observables such as:
\begin{eqnarray}
\tilde{R}_1 & \equiv &
{{\sigma}^+ - {\sigma}^- \over {\sigma}^+ + {\sigma}^-} \nonumber \\
\tilde{R}_2(y_0) & \equiv &
{ {d {\sigma}^+\over dy_\ell} \mid_{y_\ell=y_0}-
{d {\sigma}^-\over dy_\ell}\mid_{y_\ell= -y_0} \over
{d {\sigma}^+\over dy_\ell}\mid_{y_\ell=y_0} +
{d {\sigma}^-\over dy_\ell}\mid_{y_\ell= -y_0} } \nonumber \\
\tilde{R}_3(p_T) & \equiv &
{ {d {\sigma}^+\over dp_T} -{d {\sigma}^-\over dp_T} \over
{d {\sigma}^+\over dp_T} +{d {\sigma}^-\over dp_T} }
\label{cpobs},
\end{eqnarray}
where ${\sigma}^\pm$ refers to $\sigma(p\pp\ra\ell^\pm\nu X)$.
Of course, it is also possible to construct $\cp$ odd observables based
on more complicated correlations, but we will not pursue that route in this
paper.\footnote{
The capability of a detector like CDF to study asymmetries in distributions
has been demonstrated in the measurement of the charge asymmetry in
the production of $W$'s as a function of the $W$ rapidity \cite{cdf}.}

To generate the $\cp$ odd observables in Eq.~\ref{cpobs} it is necessary
to have an absorptive phase. We consider $\cp$ violating four-fermion
operators, and their one-loop
contribution to the $W f \overline{f^\prime}$ amplitudes.
The effective four-fermion operators consistent with the symmetries of the
Standard Model are listed, for example, in Ref.~\cite{buchmuller,burgess}.
Since we want to interfere the $\cp$ violating amplitude with the
lowest order standard model amplitude, we choose a four fermion operator
of the form ${\cal O}^{(3)}_{\ell q}$ in the notation of
Ref.~\cite{buchmuller}:
\begin{equation}
{\cal L}_{\cp} = {4 \pi \over \Lambda^2} e^{i\phi}\overline{c}_L\gamma_\mu s_L
\overline{\ell}_L \gamma^\mu \nu_L + {\rm ~h.~c.~}
\label{ffcp}
\end{equation}
This operator is similar to the one studied in Ref.~\cite{barbieri}, chosen
so that its interference with the Standard Model is not suppressed by powers of
light fermion masses. We keep the same normalization as Ref.~\cite{barbieri}
which considered $\ell^\pm \nu$ production away from the $W^\pm$ resonance.
We consider the operator Eq.~\ref{ffcp}, instead of a similar one with $\uu d$
quarks (used in Ref.~\cite{barbieri}) for two reasons. First, for the
operator with $\uu d$ there is a cancellation between two contributions to
$p\pp \ra \ell^\pm \nu$ as discussed in Ref.~\cite{barbieri}. This cancellation
is exact for the resonant process that we study here, but it does not occur
for the operator with $\overline{c} s$ of Eq.~\ref{ffcp}. Also, whereas
there are several indirect constraints from low energy experiments on the
operator with $\uu d$ \cite{barbieri}, analogous constraints on the operator in
Eq.~\ref{ffcp} depend on naturalness assumptions.

\begin{figure}[htp]
\centerline{\hfil\epsffile{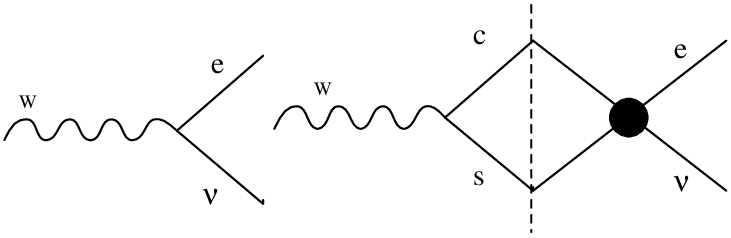}\hfil}
\caption[]{$W f \overline{f^\prime}$ vertex with $\cp$ violation and an
absorptive phase.}
\label{wff}
\end{figure}

We compute the one-loop effects of Eq.~\ref{ffcp} in the
$W f \overline{f^\prime}$ vertex as sketched in Figure~\ref{wff}.
The first diagram in Fig.~\ref{wff} is just the Standard Model
vertex and the second diagram represents the absorptive part of the
one-loop contribution from the operator in Eq.~\ref{ffcp}. This
absorptive part contains the $\cp$ violating coupling $\sin\phi$,
and can be easily computed using the Cutkosky rule. We obtain
an amplitude for $\uu d \ra e^- \nu$:
\begin{eqnarray}
{\cal M} (\uu d \ra e^-\nu) &=&  \biggl({g \over 2 \sqrt{2}}\biggr)^2
V^\star_{ud} {1\over \hat{s} -M_W^2 + i M_W \Gamma_W}
\biggl[ 1 + {V_{cs} \over 6} \sin\phi {\hat{s} \over \Lambda^2}\biggr]
\nonumber \\ &&
\overline{v}_u \gamma_\mu (1 - \gamma_5)u_d
\uu_e \gamma^\mu (1 - \gamma_5)v_\nu
\label{effver}
\end{eqnarray}
We see that the $\cp$ violating contribution to the amplitude is proportional
to $\hat{s}=(p_e+p_\nu)^2$  and is thus suppressed in pion decay.
The corresponding amplitude for $u \dd \ra e^+ \nu$ is:
\begin{eqnarray}
{\cal M} (u \dd \ra e^+\nu) &=&  \biggl({g \over 2 \sqrt{2}}\biggr)^2
V_{ud} {1\over \hat{s} -M_W^2 + i M_W \Gamma_W}
\biggl[ 1 - {V^\star_{cs} \over 6} \sin\phi {\hat{s} \over \Lambda^2}\biggr]
\nonumber \\ &&
\overline{v}_d \gamma_\mu (1 - \gamma_5)u_u
\uu_\nu \gamma^\mu (1 - \gamma_5)v_e
\label{effvert}
\end{eqnarray}

{}From these results it is clear that all the differential cross sections
can be obtained by multiplying the minimal Standard Model results by an
overall factor that depends only on $\hat{s}$. For example:
\begin{equation}
{d \hat{\sigma}^\pm \over d\ \cos\theta} =
\biggl({d \hat{\sigma}^\pm \over d\ \cos\theta} \biggr)_{SM}
\biggl[1\mp{1\over 3}{\hat{s} \over \Lambda^2}\sin\phi\biggr]
\label{cpfactor}
\end{equation}

The total hadronic cross-section is obtained as usual: integrating with the
parton distribution functions. Within our assumptions, all $\cp$ violation
occurs in the parton subprocess and the parton distribution functions satisfy
$f_{\dd/\pp}(x)=f_{d/p}(x)\equiv d(x)$ and $f_{u/\pp}(x)=f_{\uu/p}(x)\equiv
{\overline u}(x)$, etc. In this context, the asymmetries in Eq.~\ref{cpobs}
can be trivially computed using the narrow width approximation to
replace $\hat{s}$ with $M^2_W$ in the overall factor. They are:
\begin{equation}
\tilde{R}_1 = \tilde{R}_2(y_0) = \tilde{R}_3(p_T) \approx
-{1 \over 3}{M_W^2\over \Lambda^2}\sin\phi.
\label{rateas}
\end{equation}

In order to observe a signal at the one-standard deviation level,
we require that the number of events, $N$, be greater than
\beq
N >  {1\over \tilde{R}_1^2}
\approx 200,000 \biggl({\Lambda\over 1~TeV}\biggr)^4
{1\over \sin^2\phi}\quad .
\label{asy}
\eeq
The Tevatron is therefore, in principle, capable of observing $\cp$
violation coming from new physics at the $TeV$ scale with a sample
of about $10^6$ $W^\pm$ events.

We have argued that it is unlikely that $\cp$ violation in pion decays
can place a significant constraint on the strength of this operator.
Nevertheless, there are indirect constraints on the strength
of the $\cp$ conserving part of operators like Eq.~\ref{ffcp}. This
can be seen by looking at the gauge invariant version of the operator
${\cal O}^{(3)}_{\ell q}$ in the notation of Ref.\cite{buchmuller}:
\beq
{\cal L}={4\pi\over \Lambda^2}{\overline Q}_L
\tau^I \gamma_\mu Q_L {\overline L}_L \gamma^\mu \tau^I  L
\label{effopb}
\eeq
where ${\overline Q}=({\overline u},{\overline d})$
and ${\overline L}=({\overline e}, {\overline \nu}$). This contains the term:
\beq
{4\pi\over \Lambda^2} {\overline s}\gamma^\mu(1-\gamma_5)d
{\overline \nu} \gamma_\mu (1-\gamma_5) \nu
\eeq
which contributes to the decay $K^+\rightarrow \pi^+ \nu
{\overline \nu}$.  Using the present experimental bound \cite{e787},
$BR(K^+\rightarrow \pi^+\nu {\overline \nu}) < 10^{-9}$, we
obtain a limit,
\beq
\Lambda > 70-100~TeV.
\label{lims}
\eeq
Assuming that the couplings of operators involving first and second generation
fermions are of the same order, this becomes an indirect constraint on
the scale appearing in Eq.~\ref{ffcp}.

\section{$p\pp \ra W^\pm \ + \ {\rm ~1~jet}\ra \ell^\pm \nu \ + \ {\rm
{}~1~jet}$}

In this process there are several parton subprocesses that contribute
at leading order in $\alpha_S$ and there are enough independent four-vectors
to give rise to $T$-odd correlations. Ref.~\cite{nacht} has
listed several $\cp$ odd observables for this system.
We consider a few simple observables
generated by a $\cp$ violating effective $udW$ operator that respects the
symmetries of the Standard Model.

Working again in the $p\pp$ center of mass frame, a $\cp$ transformation
takes the reaction $p(\vec{p})\pp(-\vec{p}) \ra \ell^+(\vec{q}) \nu
{\rm ~jet}(\vec{p}_j) X$
into $p(\vec{p})\pp(-\vec{p}) \ra \ell^-(-\vec{q}) \nn
{\rm ~jet}(-\vec{p}_j)\overline{X}$. In this case, the $\cp$ transformation
takes all the particles that form the jet into their respective anti-particles
and reverses their momenta. To use jet variables we assume that the algorithm
that defines the
jet is CP blind in the sense that the probability of finding that a
collection of particles with certain momenta forms a jet is the same as
the probability of finding that a collection of the respective anti-particles
with the momenta reversed forms a jet \cite{donoghue}. With these
definitions, we see that the observable kinematic variables are
the rapidity and transverse momentum of the lepton and the
jet. By the same arguments of the previous section, we can construct $\cp$
odd observables identical to those in Eq.~\ref{cpobs}. In this case there
are additional distribution asymmetries obtained by replacing $y_\ell$ by
$y_{jet}$ and $p_{T\ell}$ by $p_{Tjet}$ in Eq.~\ref{cpobs}.
For example, the same $\cp$ odd interaction of Eq.~\ref{ffcp} would
generate the same asymmetries as in Eq.~\ref{rateas}.

It is also possible to have $T$-odd correlations in this process.
The interest of these correlations lies
in the fact that they can generate $\cp$ odd observables without requiring
additional absorptive phases and thus may test different types
of $\cp$ violating physics than the $T$-even asymmetries.
To construct a $\cp$ violating observable we have to compare the correlations
induced in $W^-$ plus jet production with those induced in $W^+$ plus jet
production, in the same way that we constructed $\cp$-odd asymmetries
from $T$-even observables.

For the $W$ + 1 jet process there is one $T$-odd correlation that can
be observed; in the lab frame it is given by the triple product
${\vec p}_\ell\cdot ({\vec p}_{\rm beam}\times {\vec p}_{\rm jet})$.
There are several equivalent ways to use this correlation
to construct a $T$-odd observable. The basic idea is to
define the plane formed by the beam and jet momenta and count
the number of events with the lepton above the plane minus the number of
events with the lepton below the plane:
\begin{equation}
A^\pm = \sigma^\pm  [({\vec p}_{\rm beam} \times
 {\vec p}_{\rm jet})\cdot {\vec p}_\ell > 0] -
\sigma^\pm [({\vec p}_{\rm beam} \times {\vec p}_{\rm jet})
\cdot {\vec p}_\ell < 0]
\label{tripobs}
\end{equation}
where $A^\pm$ refers to the observable for $W^\pm$ events
(or $\ell^\pm \nu$ events). A practical way to implement this observable
in the calculation (or in the experiment)
is to weigh the matrix element squared for a parton subprocess
(or to weigh the observed event) by the sign of
${\vec p}_\ell\cdot ({\vec p}_{\rm beam}\times {\vec p}_{\rm jet})$.
{}From the $\cp$ transformation for this reaction, we see
that $\cp$ symmetry predicts that $A^+=A^-$. In a manner analogous to
the $T$-even observables of Eq.~\ref{cpobs}, it is useful to construct
not only the fully integrated asymmetry, but asymmetries for distributions
as well. One obvious reason is that the simultaneous study of the
different distribution asymmetries provides a handle on the possible
$\cp$ odd biases of a detector. Another reason is that
it is possible for the integrated asymmetry to vanish while having
non-vanishing asymmetries for distributions. Some $T$-odd
$\cp$ odd observables are then:
\begin{eqnarray}
{R}_1 & \equiv &
{A^+ - A^- \over {\sigma}^+ + {\sigma}^-} \nonumber \\
{R}_2(y_0) & \equiv &
{ {d A^+\over dy} \mid_{y=y_0}-
{d A^-\over dy}\mid_{y= -y_0} \over
{d {\sigma}^+\over dy}\mid_{y=y_0} +
{d {\sigma}^-\over dy}\mid_{y= -y_0} } \nonumber \\
{R}_3(p_T) & \equiv &
{ {d A^+\over dp_T} -{d A^-\over dp_T} \over
{d {\sigma}^+\over dp_T} +{d {\sigma}^-\over dp_T} }
\label{cpobst},
\end{eqnarray}
where $y$ and $p_T$ can be the rapidity and transverse momentum
of the lepton or the jet (or the $W$). We have chosen to normalize
the asymmetries with respect to the respective differential cross-sections.
This is because non-zero but $\cp$ conserving $T$-odd correlations
arise only through final state interactions and are thus generally
small. With our normalization we obtain dimensionless asymmetries
that are not misleadingly large.

We now turn our attention to a simple effective interaction
that can generate some of these asymmetries. For $W$ plus jet production
the only parton subprocesses that can give rise to the $T$-odd correlation are
$g d \ra u W^-$, $g \uu \ra \dd W^-$ and $ \uu d \ra g W^-$.
These subprocesses receive
\begin{figure}[htp]
\centerline{\hfil\epsffile{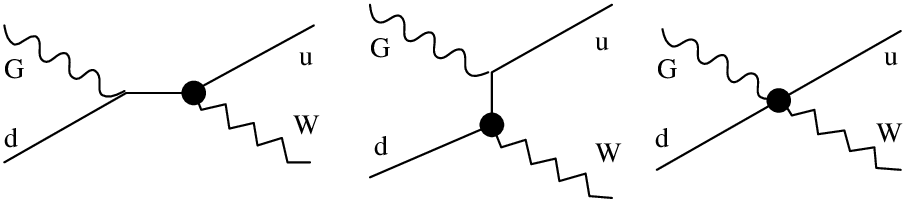}\hfil}
\caption[]{Diagrams contributing to $g d \ra u W^-$.}
\label{inig}
\end{figure}
contributions from the diagrams shown in Figure~\ref{inig}
and the crossed diagrams.
The first two diagrams in this figure correspond to the
ones occurring in the standard model with the $W$ vertex replaced
by an effective vertex that includes new physics and is represented
by the full circle. In general, this vertex will contain derivative
couplings and $SU(3)_c$ gauge invariance will
require a contact interaction as
depicted in the third diagram. Similarly, electromagnetic gauge invariance will
require a contact interaction involving a photon. The interaction
with the photon does not contribute to the process we study, but is important
for $p\pp \ra W \gamma$.

In order to generate a $\cp$-odd triple product correlation,
we need interference between the different diagrams that
give rise to each parton subprocess. This will only occur if
the diagrams have different phases.

\begin{figure}[htp]
\centerline{\hfil\epsffile{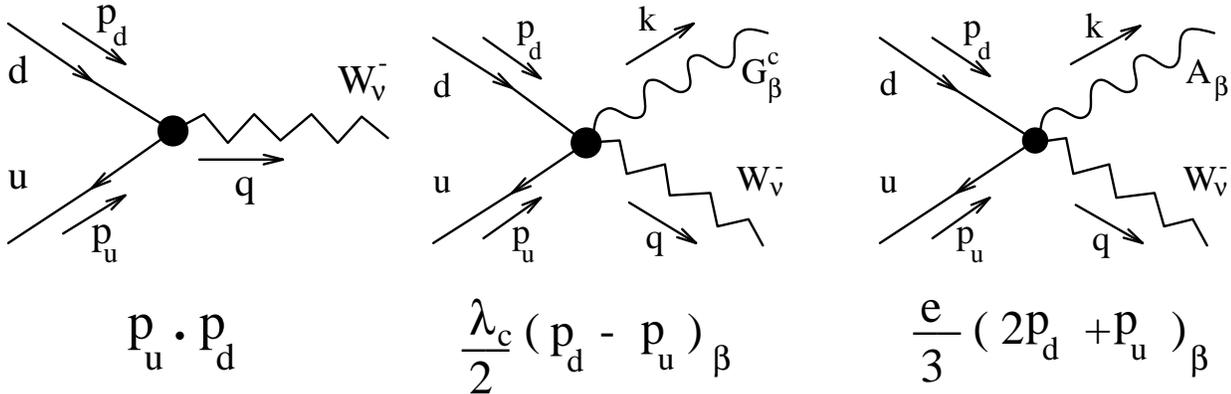}\hfil}
\caption[]{Some Feynman rules for Eq.~\ref{cplagt}, the factors in the
figure multiply $ig\tilde{\kappa}\gamma_\nu(1-\gamma_5)/(2\sqrt{2})$.}
\label{feyn}
\end{figure}

An effective $\cp$ violating $W u d$ coupling, can be written
down in the non-linear realization of electro-weak symmetry breaking as done
in Ref.~\cite{peccei}. In unitary gauge it contains the couplings:
\begin{equation}
{\cal L} = {g \over 2 \sqrt{2}}\biggl[
\kappa \uu \gamma_\mu (1 - \gamma_5) d W^{+\mu} +
\kappa^\star \dd \gamma_\mu (1-\gamma_5) u W^{-\mu}\biggr]
\label{pzlag}
\end{equation}
where there is $\cp$ violation if $\kappa$ has an imaginary part.
It is easy to see that this operator will not generate a triple product
correlation of the type we want because both diagrams\footnote{For this
operator there would not be a contact interaction as in the third diagram.} in
Figure~\ref{inig} have the same phase.
A study of the diagrams in Figure~\ref{inig} reveals that it is
possible to give them different $\cp$ phases, if the $\cp$ violation is
generated by an operator that depends on the momentum carried by the
fermions in the $W f \overline{f^\prime}$ coupling. This leads us to consider
a higher dimension operator, similar to Eq.~\ref{pzlag}, of the form:
\begin{equation}
{\cal L}= -{\sqrt{2}\over\Lambda^2}
\biggl[\tilde{\kappa}
\overline{\Psi}_L \overleftarrow{D}_\alpha
\gamma_\mu \Sigma \tau_- \Sigma^\dagger
\overrightarrow{D}^\alpha  \Psi_L \Sigma^\mu_+ +
\tilde{\kappa}^\star \overline{\Psi}_L  \overleftarrow{D}_\alpha
\gamma_\mu \Sigma \tau_+ \Sigma^\dagger
\overrightarrow{D}^\alpha  \Psi_L \Sigma^\mu_- \biggr].
\label{cplagt}
\end{equation}
The notation is the same as that in Ref.~\cite{peccei}: in unitary gauge
$\Sigma=1$ and
$\Sigma^\mu_\pm = -{g\over 2}  W^{\mu \pm}$. For the processes
of interest there
will only be one $W$ boson and no $Z$ bosons, so the covariant derivatives
refer only to QED and QCD:
\begin{equation}
D_\alpha \Psi_L \ra (\partial_\alpha + {i \over 2}g_S\lambda^a G^a_\alpha
+ i e QA_\alpha) \biggl(\begin{array}{c} u \\ d_\theta \end{array}\biggl)_L.
\label{covd}
\end{equation}
We present some of the Feynman rules from Eq.~\ref{cplagt}
in Figure~\ref{feyn}.
{}From these Feynman rules it is clear that this operator will give different
$\cp$ phases to the diagrams in Fig.~\ref{inig} and will induce
a $\cp$ violating triple product correlation.

For the gluon initiated process $g d \ra u e^- \nu$
we find that the interference
of the new interaction with the lowest order standard model amplitude generates
the contribution to the spin and color averaged matrix element squared,
(within the narrow width approximation):
\footnote{$\epsilon(p_1,p_2,p_3,p_4)\equiv \epsilon_{\alpha\beta\gamma\delta}
p_1^\alpha p_2^\beta p_3^\gamma p_4^\delta$.}
\beq
 \sum\biggl| {\cal M}\biggl(g(p_g)d(p_d)\rightarrow
u(p_u) e^-(p_e) {\overline \nu}(p_\nu)\biggr)\biggr|^2=
{C\over 24} M(p_g,p_d,p_u,p_e,p_\nu)
\label{interft}
\eeq
where
\beq
C=-{(4 \pi)^4\over M_W\Gamma_W}
{\alpha^2 \alpha_s\over s_\theta^4} {Im ~{\tilde \kappa}\over 2 \Lambda^2}
\delta(p_W^2-M_W^2),
\eeq
 $p_W^2=2 p_e\cdot p_\nu$, $\Gamma_W$ is the total $W$ decay width
and
\beq
M(p_1,p_2,p_3,p_e,p_\nu)\equiv \epsilon(p_1,p_2,p_3,p_e)
\biggl[ {p_2\cdot(p_\nu-p_e)\over p_2\cdot p_1}
+{p_3\cdot (p_\nu-p_e)\over p_3\cdot p_1}
\biggr]\quad .
\eeq
We ignore interference terms which do not contribute to the $\cp$ violating
observables in Eq.~\ref{cpobst}. For our estimates we use for the
matrix element squared for this process the sum of the
standard model contribution and  Eq.~\ref{interft}.
We take the convention that $p_1,p_2$ are incoming and
$p_3,p_e,p_\nu$ are outgoing.  The other parton level
sub-process amplitudes can be found using crossing symmetry:

\beqn
A^{g {\overline d}}(p_g,p_d,p_u,p_e,p_\nu)&=&
{C\over 24}M(p_g,p_d,p_u,p_e,p_\nu)
\nonumber \\
A^{g {\overline u}}(p_g,p_u,p_d,p_e,p_\nu)&=&
{C\over 24}M(p_g,-p_d,-p_u,p_e,p_\nu)
\nonumber \\
A^{g u}(p_g,p_u,p_d,p_e,p_\nu)&=&
{C\over 24}M(p_g,-p_d,-p_u,p_e,p_\nu)
\nonumber \\
A^{{\overline u} d}(p_u,p_d,p_g,p_e,p_\nu)&=&
-{C\over 9}M(-p_g,p_d,-p_u,p_e,p_\nu)
\nonumber \\
A^{u {\overline d}}(p_d,p_u,p_g,p_e,p_\nu)&=&
-{C\over 9}M(-p_g,p_d,-p_u,p_e,p_\nu)\quad .
\eeqn

We have done a numerical simulation to estimate the size of the
asymmetries of Eq.~\ref{cpobst} induced by the interaction Eq.~\ref{cplagt}.
We impose a cut on the jet rapidity that simulates the typical acceptance
at the Tevatron $|y_{jet}|< 3.$ For illustration purposes we choose
the cut $p_{Tjet}> 30$~GeV to define the jet.
Increasing the value of this cut reduces the total
number of events but increases the signal to background ratio because
the $\cp$ odd contribution is not peaked at low $p_T$.

We present our results for Im~$\kappa =1$ and $\Lambda=1$~TeV in
Eq.~\ref{cplagt}, and remind the reader that they scale as
Im~$\kappa/\Lambda^2$.

We find that the fully integrated asymmetry ${R}_1$ vanishes for
the interaction Eq.~\ref{cplagt}. To our knowledge, there is no reason
for this asymmetry to vanish in general, so this result is specific to
the example we chose.

\begin{figure}[htp]
\centerline{\hfil\epsffile{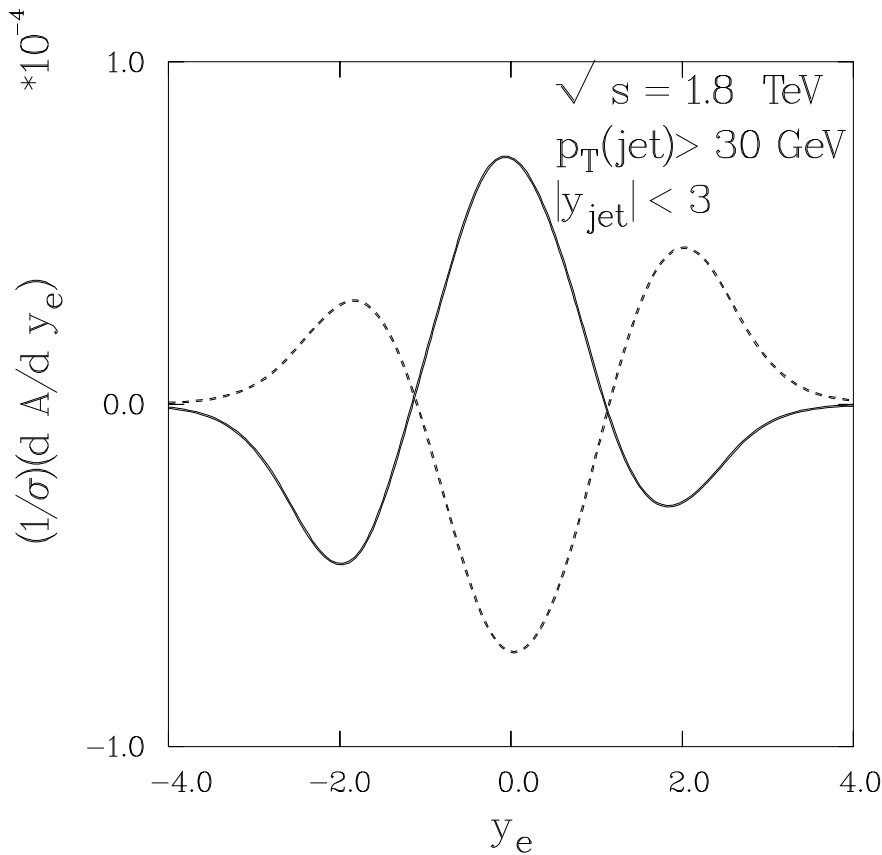}\hfil}
\caption[]{Asymmetries $dA^-/dy_e$ (solid line) and $dA^+/dy_e$ (dotted line)
normalized to the lowest order Standard Model cross section $\sigma (p \pp \ra
e^\pm \nu {\rm jet})$.}
\label{dady}
\end{figure}

In Figure~\ref{dady} we present the distributions of the observable
Eq.~\ref{tripobs} with respect to the lepton rapidity generated by the
$\cp$ violating interference Eq.~\ref{interft}. The corresponding
($\cp$ conserving)
distributions for the Standard Model vanish to the order we work (they
are non-zero at higher order in QCD\cite{hagi}).
{}From that figure we can see
that although the distributions do not vanish as a function of the lepton
rapidity, the integrated observable does vanish. As explained before,
a non-zero value of these distributions is not a signal of $\cp$
violation. Rather, the signal of $\cp$ violation is the fact that
the distribution for $W^+$ at a given value of $y_e$ is not equal to
the distribution for $W^-$ at the value $-y_e$.

\begin{figure}[htp]
\centerline{\hfil\epsffile{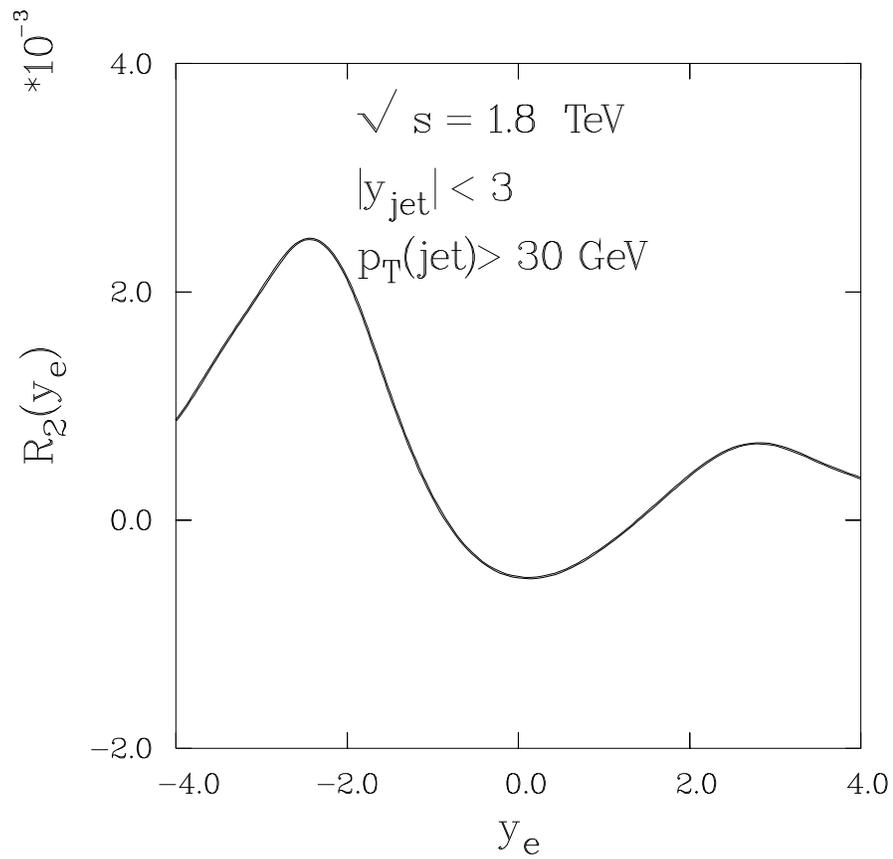}\hfil}
\caption[]{$\cp$ violating asymmetry in the lepton rapidity distributions
as defined in Eq.~\ref{cpobst}. We used
$\Lambda = 1$~TeV and Im~$\kappa =1$ in Eq.~\ref{cplagt}}
\label{yasym}
\end{figure}

In Fig.~\ref{yasym} we present the asymmetry in
the lepton rapidity distribution ${R}_2(y_\ell)$.
At the one standard deviation level
some $10^6$ $W^\pm$ plus one jet events would be needed to observe this
asymmetry. Measurement of this asymmetry for arbitrary values
of $y_e$ is complicated by the fact that the acceptance of the
detector must be the same for $y_e$ and $-y_e$. Figure~\ref{yasym} shows
that the asymmetry doesn't necessarily vanish at $y_e=0$, making this
a particularly interesting point to search for $\cp$ violation.

\begin{figure}[htp]
\centerline{\hfil\epsffile{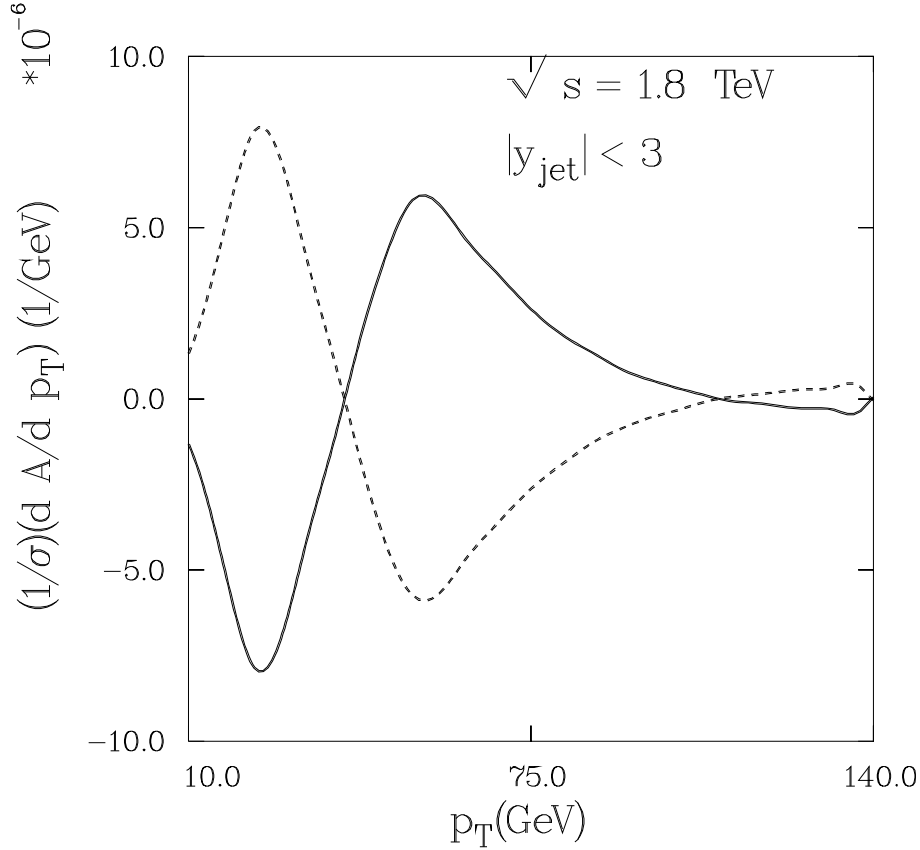}\hfil}
\caption[]{Asymmetries $dA^-/dp_{Te}$ (solid line) and $dA^+/dp_{Te}$
(dotted line)
normalized to the lowest order Standard Model cross section $\sigma (p \pp \ra
e^\pm \nu {\rm jet})$}
\label{ptedis}
\end{figure}

\begin{figure}[htp]
\centerline{\hfil\epsffile{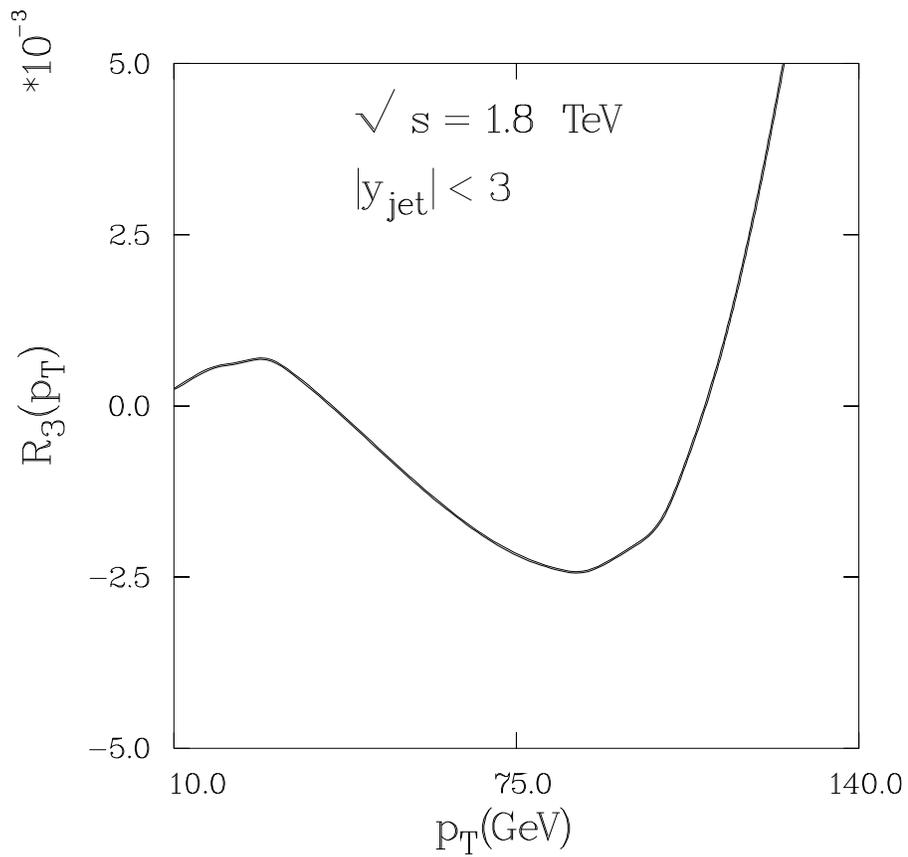}\hfil}
\caption[]{$\cp$ violating asymmetry in the lepton transverse momentum
distributions as defined in Eq.~\ref{cpobst}. We used
$\Lambda = 1$~TeV and Im~$\kappa =1$ in Eq.~\ref{cplagt}.}
\label{ptasym}
\end{figure}

We show in Figure~\ref{ptedis} the distributions of the $A^\pm$
correlations with respect to the lepton transverse momentum. The
$\cp$ violating nature of the interaction is reflected in the fact that
the two distributions have opposite signs. Normalizing as in Eq.\ref{cpobst}
we present $R_3(p_{Te})$ in Figure~\ref{ptasym}.
The rise in the asymmetry for large values of $p_{Te}$ is due to the
decrease in the standard model distribution $d\sigma/dp_{Te}$ which is
peaked around $M_W/2$. This increase in the asymmetry is thus
accompanied by a decrease in the total number of events.

\section{Conclusions}

We have constructed several $\cp$-odd asymmetries that can be used
to search for $\cp$ violation in $W^\pm + (0,1)~{\rm jet}$ events in
$p\pp$ colliders. We have estimated the contributions to these
asymmetries from some simple $\cp$ violating effective operators that
respect the symmetries of the Standard Model. Assuming that the scale
of the new physics responsible for these operators is 1 TeV, we find that
it is possible to search for $\cp$ violation at the Tevatron with as few
as $10^6$ events. Similar observables can be constructed for other
processes such as $p\pp \ra W^\pm \gamma$.

\vspace{1in}

\noindent {\bf Acknowledgements} The work of G.V. was supported in
part by a DOE OJI award under contract number DEFG0292ER40730. G. V. thanks
the theory group at BNL for their hospitality while part of this work
was performed. We are grateful to S. Errede, T. Han, J. Hauptman
and S. Willenbrock for helpful discussions.

\end{document}